\documentclass[preprint,5p,twocolumn]{elsarticle}

\usepackage{booktabs} 
\usepackage{hyperref}
\usepackage{overpic}
\usepackage{tikz}
\usetikzlibrary{positioning}

\usepackage{amsmath,amsfonts}
\usepackage{algorithmic}
\usepackage{graphicx}
\usepackage{xspace}
\usepackage{textcomp}
\usepackage{subcaption}
\usepackage{siunitx}
\usepackage{multirow}
\usepackage{subfiles}
\usepackage{xspace}
\usepackage{svg}
\usepackage{tablefootnote}
\usepackage{pifont}
\usepackage{url}
\usepackage{soul}
\usepackage{threeparttable}

\usepackage[linesnumbered,ruled,vlined]{algorithm2e}
\SetKwComment{Comment}{\{}{\}}

\usepackage[T1]{fontenc}

\usepackage{mathtools}
\usepackage{soul}
\usepackage{comment}

\definecolor{boristext}{rgb}{0.22, 0.44, 0.88}
\definecolor{migueltext}{rgb}{0.49, 0.37, 0.56}
\definecolor{boriscomments}{rgb}{0.88, 0.04, 0.04}
\definecolor{miguelcomments}{rgb}{0.5, 0, 0.8}
\definecolor{boristochange}{rgb}{0.2, 0.8, 0.8}
\definecolor{ferrantext}{rgb}{0.37, 0.54, 0.56} 
\definecolor{ferrancomments}{rgb}{0, 0.8, 0.8} 
\usepackage{tabulary}

\definecolor{boristext}{rgb}{0.22, 0.44, 0.88}
\definecolor{boriscomments}{rgb}{0.88, 0.04, 0.04}
\definecolor{boristochange}{rgb}{0.2, 0.8, 0.8}

\begin{document}

\begin{frontmatter}
\title{NeSt-VR: An Adaptive Bitrate Algorithm for Wireless Virtual Reality Streaming}

\journal{Computer Networks}

\author[upf]{Miguel Casasnovas\corref{equal}}
\author[upf]{Ferran Maura\corref{equal}}
\author[crew]{Isjtar Vandebroeck}
\author[crew]{Haryo Sukmawanto}
\author[crew]{Eric Joris}
\author[upf]{Boris Bellalta}

\cortext[equal]{Corresponding authors}
\address[upf]{Universitat Pompeu Fabra, Barcelona, Spain \\ Email: {\{firstname.lastname\}@upf.edu}\vspace{1.5ex}}
\address[crew]{CREW, Brussels, Belgium \\ Email: {lab@crew.brussels} \vspace{-2ex}}

\begin{abstract} 
Interactive Virtual Reality (VR) streaming over wireless links requires both high frame rates and low motion‑to‑photon latency, requirements that are difficult to sustain under changing channel conditions such as fluctuating bandwidth and multi‑user contention. Adaptive BitRate (ABR) control is therefore essential, as it dynamically adjusts the encoded bitrate in response to network conditions to maintain smooth interactive video delivery. In this paper, we present NeSt‑VR, the \textit{Network‑aware Step‑wise ABR algorithm for VR streaming}, a configurable controller that uses application‑level frame‑delivery and frame‑delay feedback to adjust the encoder target bitrate in discrete steps, reducing abrupt bitrate oscillations while supporting a satisfactory interactive VR user experience. We evaluate NeSt‑VR through Wi‑Fi experiments against state‑of‑the‑art ABR baselines, including GCC, NADA, and EVeREst-Intra. The evaluation covers single‑user and multi‑user scenarios and common Wi‑Fi challenges such as co‑channel interference and capacity fluctuations. The results show that NeSt‑VR effectively manages bandwidth variation, maintains frame delivery and low latency, and compares favorably with the baseline controllers in the evaluated scenarios.
\end{abstract} 


\end{frontmatter}



\section{Introduction}

Virtual, Mixed, and Augmented Reality (VR, MR, AR), technologies encompassed by the term eXtended Reality~(XR), have increased in popularity over the last decade~\cite{minopoulos2022opportunities}, driven by the widespread availability of consumer XR head-mounted displays~(HMDs) such as the HTC Vive, Valve Index, and Meta Quest. Since \textit{local rendering} (performing rendering directly on the user's device) is limited by the device's computational capabilities, \textit{remote rendering}~\cite{shi2015survey} (offloading the computational workload to powerful cloud or edge computing servers) is often required, particularly for applications with high graphical demands.

Traditionally, remote rendering has relied on wired connections to ensure low-latency, high-bandwidth transmissions. However, wireless technologies such as 5G and Wi-Fi provide greater flexibility, especially for mobile XR applications, as they enable users to move freely without being tethered to a physical connection~\cite{akyildiz2022wireless}. Among these technologies, Wi-Fi is expected to become the most prevalent technology for XR streaming due to its widespread adoption in home and office spaces and the anticipated improvements in throughput, reliability, and latency in future IEEE 802.11 standard revisions~\cite{galati2024will,adame2021time}.

However, Wi-Fi networks operate in unlicensed frequency bands and are susceptible to external perturbations such as co-channel interference (overlapping transmissions on the same frequency band)~\cite{deek2013intelligent}, and path loss (signal degradation due to distance from the access point or obstructions)~\cite{adame2019tmb}. In addition, Wi-Fi is prone to channel contention (competition for access to the wireless medium) and network congestion (demand exceeding available bandwidth), particularly in high-density environments~\cite{barrachina2021wi}. These factors can increase latency, jitter, and packet loss, leading to visual artifacts, stuttering, and frame drops that degrade the user's quality of experience in VR streaming~\cite{itutg1035}.

To mitigate congestion episodes that could otherwise disrupt VR streaming sessions, Adaptive BitRate~(ABR) control is a natural solution. ABR algorithms adjust the encoder target bitrate based on observed network conditions, reducing the bitrate when congestion is detected or anticipated to alleviate pressure on the network, and increasing it when more capacity becomes available to improve rendered video quality. Traditional ABR schemes, such as FESTIVE~\cite{jiang2012improving} and BOLA~\cite{spiteri2020bola}, were originally designed for on‑demand video streaming, where sizeable playback buffers can absorb temporary stalls and quality fluctuations. These design assumptions do not directly translate to interactive VR streaming, which typically relies on shallow buffering and must respect motion‑to‑photon latency budgets on the order of tens of milliseconds. Consequently, specialized ABR strategies that explicitly account for interactive VR characteristics are needed, as conventional mechanisms may be less effective unless adapted.

In this paper, we present NeSt‑VR, the \emph{Network‑aware Step‑wise ABR algorithm for VR streaming}, a configurable controller for interactive VR streaming over Wi‑Fi. NeSt‑VR uses application‑level frame‑delivery and frame‑delay feedback to adjust the encoder target bitrate in discrete steps, supporting a satisfactory interactive VR user experience while avoiding persistent overload or underutilization of the available network capacity. We implement NeSt‑VR on top of Air Light VR (ALVR)\footnote{\url{https://github.com/alvr-org/ALVR}}, an open‑source VR streaming platform, and evaluate it using Wi‑Fi real‑world experiments against state‑of‑the‑art non‑VR‑specific and VR‑specific ABR schemes.

In particular, the contributions of this paper are:
\vspace{-0.03in}
\begin{enumerate}
    \item Identifying practical operating regions for frame delivery and end‑to‑end frame delay that correlate with satisfactory VR streaming user experiences can guide the design and evaluation of VR‑aware ABR controllers.
    \item Proposing NeSt‑VR, a VR‑aware step‑wise ABR algorithm that uses frame‑delivery and frame round‑trip‑time measurements from the VR application to drive bitrate adaptations, reducing abrupt bitrate oscillations while preserving interactive streaming quality.
    \item Comparing NeSt‑VR against ALVR's native adaptive mode and recent ABR baselines, including GCC~\cite{carlucci2016analysis}, NADA~\cite{zhu2020network}, and EVeREst~\cite{liubogoshchev2021everest, korneev2024model}, across diverse experimental scenarios.
    \item Integrating NeSt‑VR and the considered ABR baselines within ALVR, and releasing our ALVR v20.6.0 fork as open‑source on GitHub\footnote{\label{fn:github}\url{https://github.com/wn-upf/NeSt-VR/tree/abr_comparison}} to facilitate reproducible evaluation and future extensions.
    \end{enumerate}

This work extends our previous study~\cite{maura2024experimenting} by refining the NeSt-VR design with separate bitrate increase and decrease logic, parameter recommendations, and predefined profiles. It also broadens the evaluation to state‑of‑the‑art ABRs and more demanding Wi‑Fi conditions, including multi‑user contention and co‑channel interference, and validates NeSt‑VR in a professional out‑of‑the‑lab deployment.


\section{Related Work}\label{sec:relatedwork_abr}
Adaptive bitrate control has been extensively studied in conventional video streaming and real-time media systems. DASH-based methods such as BOLA~\cite{spiteri2020bola} and transport-oriented WebRTC controllers such as GCC~\cite{carlucci2016analysis} estimate network conditions from client-side or transport-level feedback and adapt the video rate accordingly. However, these schemes were designed for conventional video delivery and interactive media rather than remote VR streaming, where very high bitrates, strict delay constraints, and sensitivity to frame loss make bitrate control substantially more challenging. Consequently, remote VR streaming requires dedicated bitrate adaptation mechanisms that account for its tighter latency and reliability requirements.

Several works have therefore proposed bitrate adaptation mechanisms specifically for VR streaming. Vergados~\textit{et al.}~\cite{vergados2023adaptive} present a fuzzy-logic controller that adjusts the encoding bitrate according to the server transmission buffer occupancy for VR delivery over TCP. EVeREst~\cite{liubogoshchev2021everest} and its later extension EVeREst-Intra~\cite{korneev2024model} estimate congestion from frame-delivery delay and frame-size-based throughput measurements to derive conservative bitrate decisions under varying network load while reserving capacity for additional users. Their distributed client-side design enables adaptation without requiring network support, and experimental results show substantial reductions in frame loss and improvements in overall system goodput compared with BOLA and other DASH-style adaptation schemes in multi-user VR streaming scenarios.

Similarly, Alhilal \textit{et al.}~\cite{alhilal2025congestion} evaluate general-purpose real-time media congestion controllers for VR streaming. Specifically, they integrate GCC and NADA~\cite{zhu2020network} into ALVR and compare them with ALVR's native adaptive mode in a VR cloud-gaming testbed. Their results show that GCC's delay-based and conservative rate selection achieves substantially lower frame latency and fewer congestion-induced frame drops than either NADA or ALVR's native controller, albeit at the expense of lower delivered image quality; in contrast, NADA tends to select more aggressive bitrates and provides better bandwidth fairness when competing with loss-based flows, but can exceed the actually sustainable transmission rate.

Beyond bitrate adaptation, several studies have considered multi-dimensional adaptation strategies. Lee \textit{et al.}~\cite{lee2025adaptive} propose a quality of experience-driven adaptation algorithm for ALVR that jointly adjusts frame rate, rendering resolution, and bitrate using real-time streaming metrics to maximize user-perceived quality in VR gaming. In contrast, Sun \textit{et al.}~\cite{sun2022enabling} investigate VR streaming over 5G networks using a deep reinforcement learning framework that adapts both application- and network-level parameters based on streaming and radio measurements, but does not target bitrate adaptation directly.

Complementarily, other studies focus on spatial quality adaptation rather than stream-level bitrate control, emphasizing human visual perception through gaze-aware and foveated streaming. FovOptix~\cite{alhilal2024fovoptix} combines foveated rendering with adaptive foveated video encoding to deliver higher visual quality within the user's region of interest while reducing bitrate and computational cost in the peripheral field of view, using a GCC-inspired delay-based congestion controller to react to network variations. EyeNexus~\cite{wu2025eyenexus} extends this concept by combining real-time gaze-driven video encoding with adaptive adjustment of the foveation region according to both gaze behavior and network conditions, thereby coupling perceptual quality adaptation with bandwidth-aware control.


Despite the growing interest in VR-specific bitrate adaptation, only a few dedicated approaches have been proposed, and their evaluation has largely been limited to simplified deployment scenarios. To address this gap, this paper presents NeSt-VR, originally introduced in~\cite{maura2024experimenting}, a bitrate controller designed to meet the stringent delay and reliability requirements of interactive VR streaming, and evaluates it experimentally against both general-purpose real-time congestion controllers and VR-specific bitrate adaptation mechanisms under challenging wireless conditions.


\begin{figure}[t]
    \centering
    \includegraphics[width=\columnwidth]{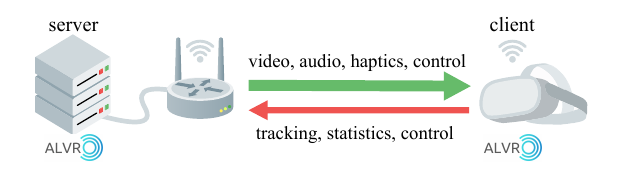}
    \caption{ALVR streaming process.}
    \label{Fig:VRStreamingPipeline}
\end{figure}

\section{Remote VR Streaming using ALVR}
Remote VR streaming enables the delivery of immersive VR experiences by offloading rendering from a resource-constrained HMD to a powerful remote server. The server renders the VR scene and streams the resulting content, including encoded video frames, to the untethered HMD over a communication network. This architecture enables high-fidelity graphics without requiring substantial computational resources on the HMD. However, it also imposes stringent requirements on network throughput, end-to-end latency, jitter, and reliability to maintain a responsive and comfortable user experience. Wireless technologies, particularly Wi-Fi, are well suited for remote VR streaming because they preserve user mobility, although their time-varying capacity, interference, and contention can significantly affect streaming performance.

ALVR is an open-source remote VR streaming platform that interfaces with SteamVR\footnote{\url{https://store.steampowered.com/app/250820/SteamVR}} to stream VR content over Wi-Fi. As illustrated in Fig.~\ref{Fig:VRStreamingPipeline}, the server---acting as the streamer---sends encoded video, audio, control information, and haptic feedback to the HMD over the downlink to the HMD. In contrast, the HMD---functioning as the client---transmits tracking data, user inputs, and runtime statistics to the server over the uplink, enabling real-time interaction and adaptive streaming.

Specifically, the client acquires head and controller tracking data from its sensors and transmits it to the server. The server uses this information for pose prediction in SteamVR, executes the VR game logic, and renders the graphical layers. These layers are composited into a video frame~(VF), encoded, fragmented into multiple packets, and transmitted to the client. Upon reception, the client reassembles and decodes the video frame before submitting it to the VR runtime for display. Once the video frame is presented, the client reports frame-related runtime statistics to the server for logging.

\begin{figure}[t] 
     \centering
     \includegraphics[width=\columnwidth]{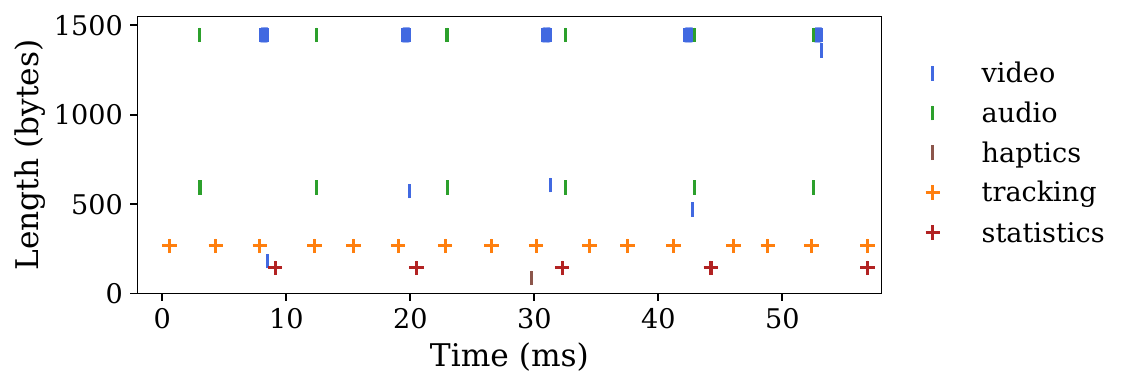}
     \caption{ALVR v20.6.0 traffic using CBR at 100 Mbps, 90 FPS, UDP, and SteamVR Home. Derived from parsed Wireshark traffic traces. Packet length includes transport headers.}
     \label{Fig:ALVR_traffic}
     \end{figure}

Fig.~\ref{Fig:ALVR_traffic} illustrates the resulting network traffic. In the downlink, each video frame is transmitted as a burst every $1/\mathrm{FPS}$ (e.g., every 11.1~ms at 90~FPS). Audio data is transmitted every 10~ms. 
Haptic feedback is transmitted opportunistically 
depending on user interactions and application events. In the uplink, tracking data is transmitted three times per frame, 
while statistics are sent once per frame, after the corresponding video frame is submitted to the VR runtime. 

\section{NeSt-VR}\label{Sec:heuristic}
NeSt-VR is a configurable ABR controller for interactive VR streaming over wireless networks. It periodically adjusts the encoder target bitrate according to frame-delivery reliability and frame delay to maintain acceptable streaming quality while avoiding bitrate oscillations and persistent network congestion.

\subsection{Control inputs}

NeSt-VR relies on two smoothed control inputs: the \emph{network frame ratio}~(NFR) and the \emph{video frame round-trip time}~(VF-RTT).
The NFR measures frame-delivery reliability over a control interval as the ratio between successfully received video frames and transmitted video frames.
The VF-RTT measures the elapsed time between transmission of a video frame by the server and reception of the corresponding runtime statistics from the client. Consequently, it includes both the downlink transmission of the video frame and the uplink transmission of its feedback. Together, these application-level quality-of-service indicators provide insight into frame losses and end-to-end delay, two critical factors influencing user-perceived quality of experience~\cite{itutg1035}.

To reduce sensitivity to transient fluctuations, both signals are smoothed before entering the control loop. The smoothing methods supported are: (i) a sliding window over the last $n$ samples, (ii) a sliding window spanning $t$ seconds, and (iii) an exponentially weighted moving average with weight $\omega$.

\subsection{Step-wise control}
As described in Algorithm~\ref{alg:nestvr}, NeSt-VR updates the target bitrate every $\tau$ seconds from a discrete ladder rather than treating it as a continuous variable. The ladder is defined by the minimum bitrate $B_{\min}$, the maximum bitrate $B_{\max}$, and the number of intermediate steps $N_{\text{s}}$.

At each adjustment period, the controller evaluates the smoothed measurements according to a hierarchical decision logic:
\begin{itemize}
    \item If the averaged NFR falls below the reliability threshold $\rho$, the controller decreases the target bitrate by $N_{\text{dw}}$ ladder steps.
    \item If the averaged NFR is acceptable but the averaged VF-RTT exceeds the delay threshold $\sigma$, the controller probabilistically decreases the bitrate: with probability $\gamma_{\text{rtt}}$ it reduces the bitrate by $N_{\text{dw}}$ steps; otherwise, it keeps the bitrate unchanged.\footnote{$\gamma_{\text{rtt}}$ controls how aggressively it reacts to excessive delays under acceptable reliability.} 
    \item If both signals are within their respective thresholds, the network may be capable of sustaining a higher bitrate. Accordingly, with probability $\gamma_{\text{+}}$, the controller increases the bitrate by $N_{\text{up}}$ ladder steps; otherwise, it keeps the bitrate unchanged.\footnote{$\gamma_{+}$ controls cautious exploration of higher bitrates under stable conditions, avoiding abrupt increases and reducing the likelihood of synchronized bitrate changes among competing streams.}
    
\end{itemize}

After each decision, the bitrate is clamped to the estimated-capacity ceiling $m \cdot C$, where $C$ is obtained from a smoothed estimate of the peak throughput observed during recent frame bursts. This ensures that the controller operates conservatively and does not exceed the fraction $m$ of the estimated capacity $C$. 
\\

Overall, NeSt-VR combines reliability-driven bitrate reduction, delay-aware bitrate adaptation, and conservative bitrate probing. Its step-wise design limits abrupt quality variations that could negatively affect the user's sense of presence and immersion.

\subsection{Safeguard for Missing Feedback}

NeSt-VR relies on per-frame feedback to estimate network conditions. Consequently, prolonged feedback loss may delay detection of network degradation and cause the controller to operate above the sustainable bitrate. To mitigate this situation, NeSt-VR activates a safeguard after an initial warm-up period $t_\text{warm}$. If no feedback update is received for longer than $k_{\text{miss}}$ frame intervals, the controller immediately performs a bitrate reduction twice as large as the standard downshift. 
This mechanism avoids unnecessary reactions during startup while providing rapid recovery from persistent feedback loss.

\subsection{Parameterization and Profiles}

NeSt-VR defines a set of configurable parameters that govern its behavior under different network conditions and application requirements (see Table~\ref{tab:heur-values} for the parameter list and Table~\ref{tab:recommendations} for the recommended values, whose rationale is discussed in Section~\ref{sec:nestvr_settings}). It also provides three predefined operating profiles that differ only in the bitrate reduction strategy while preserving the same control logic. The \textit{Balanced} profile uses symmetric adaptation ($N_{\text{dw}} = N_{\text{up}}$), the \textit{Speedy} profile performs more aggressive reductions by setting $N_{\text{dw}} = 2N_{\text{up}}$, and the \textit{Anxious} profile immediately reduces the bitrate to the minimum ladder level upon congestion ($N_{\text{dw}} = N_{\text{s}}$). These profiles offer different trade-offs between responsiveness and visual stability.

\begin{table}[t!!]
    \centering
    \setlength{\tabcolsep}{3pt}
    \caption{NeSt-VR parameters.}
    \scriptsize
    \begin{tabular}{@{}ll@{\hspace{4pt}}|@{\hspace{4pt}}ll@{}}
    \toprule
    Adjustment interval (s) & $\tau$ &  Averaging param. & $n$/$t$/$\omega$ \\
    Est. capacity multiplier & $m$ & min, max Bitrate & $B_{\min}$, $B_{\max}$  \\
    VF-RTT thresh. (ms) & $\sigma$ & initial Bitrate & $B_{\text{init}}$  \\
    NFR thresh. & $\rho$   & Bitrate steps count & $N_{\text{s}} $\\
    VF-RTT adj. prob. & $\gamma_{\text{rtt}}$  & Bitrate inc. steps & $N_{\text{up}}$\\
    Bitrate inc. explor. prob. & $\gamma_{\text{+}}$ & Bitrate dec. steps & $N_{\text{dw}}$\\
    Missing-feedback window & $k_{\text{miss}}$ & Warm-up period & $t_{\text{warm}}$ \\
    \bottomrule 
    \end{tabular}
    \label{tab:heur-values}
\end{table}

\begin{table}[t]
    \centering
    \caption{Recommended NeSt-VR parameter values used throughout the paper.}    \label{tab:recommendations}
    \scriptsize
    \begin{threeparttable}
        \begin{tabular}{@{}ll|ll|ll|ll|ll|ll@{}}
            \toprule
            $\tau$ & 1 & $\sigma$ & 22 &  $\gamma_{\text{rtt}}$ & 1 & $m$ & 0.9 & $N_{\text{up}}$ & 1 & $t_{\mathrm{warm}}$ & 5~s \\
            $t$ & 1& $\rho$ & 0.99 & $\gamma_{\text{+}}$ & 0.25 & $N_{\text{s}}$ & 9 & $N_{\text{dw}}$ & 1 & $k_{\mathrm{miss}}$ & 5 \\
            \bottomrule
        \end{tabular}
    \begin{tablenotes}[flushleft]
    \item Recommended values correspond to the Balanced profile ($N_{\text{dw}}=N_{\text{up}}$). 
    \item $B_{\min}$, $B_{\max}$, and $B_{\text{init}}$ depend on the application's bitrate requirements. 
    \end{tablenotes}
    \end{threeparttable}
\end{table}

{\renewcommand{\baselinestretch}{1.2}
\begin{algorithm}[t]
\caption{NeSt-VR adaptive bitrate controller}
\label{alg:nestvr}
\scriptsize
\KwIn{Controller parameters and runtime feedback}
\KwOut{Target bitrate values $\{B_k\}$}

$\Delta B \gets (B_{\max} - B_{\min}) / N_{\text{s}}$\;
$\mathcal{B} \gets \{B_{\min}, B_{\min} + \Delta B, B_{\min} + 2\Delta B, \dots, B_{\max}\}$\;
$B_0 \gets \max\{B \in \mathcal{B} \mid B \leq \max(B_{\text{init}}, B_{\min})\}$\;

\For{$k \gets 1$ \KwTo $\infty$}{
    \uIf{$k \geq t_{\text{warm}}/\tau$ \textbf{and} feedback timeout exceeds $k_{\text{miss}}$ frame intervals}{
            $B_k \gets \max(B_{\min}, B_{k-1} - 2 N_{\text{dw}}\cdot \Delta B)$ \Comment*[r]{-}
    }
     \Else{
    Compute \textit{avg.} NFR, \textit{avg.} VF-RTT, and capacity $C$\;
    $B_{\text{cap}} \gets \max\{B\in\mathcal{B}\mid B\le m\cdot C\}$\;

    \uIf{\textit{avg.} NFR $< \rho$}{
        $B_k \gets \max(B_{\min}, B_{k-1} - N_{\text{dw}}\cdot \Delta B)$ \Comment*[r]{-}
    }
    \uElseIf{\textit{avg.} VF-RTT $> \sigma$}{
        Sample $r_{\text{rtt}} \sim \mathcal{U}(0, 1)$\;
        \uIf{$r_{\text{rtt}} \leq \gamma_{\text{rtt}}$}{
            $B_k \gets \max(B_{\min}, B_{k-1} - N_{\text{dw}}\cdot \Delta B)$ \Comment*[r]{-}
        }
        \Else{
            $B_k \gets B_{k-1}$ \Comment*[r]{=}
        }
    }
    \Else{
        Sample $r_{\text{+}} \sim \mathcal{U}(0, 1)$\;
        \uIf{$r_{+} \leq \gamma_{+}$}{
            $B_k \gets \min(B_{\max}, B_{k-1} + N_{\text{up}}\cdot \Delta B)$ \Comment*[r]{+}
        }
        \Else{
            $B_k \gets B_{k-1}$ \Comment*[r]{=}
        }
    }
    }
    $B_k \gets \min(B_k,B_{\text{cap}})$\;
    Output $B_k$\;
}
\end{algorithm}}


\section{Baseline Bitrate Controllers}\label{sec:controllers}
We compare NeSt‑VR against several bitrate controllers that serve as baseline configurations in our evaluation: a Constant BitRate (CBR) baseline, ALVR's native adaptive mode (\textit{ALVR ABR}), two general real‑time media congestion controllers (GCC and NADA), and one cloud‑VR‑specific bitrate adaptation scheme (EVeREst-Intra). The following subsections briefly summarize each controller.

\subsection{Constant Bitrate}
CBR maintains the encoder target bitrate at a fixed value selected before the session starts. In principle, this yields a consistent encoding configuration and nominal video quality as long as the network can sustain the configured bitrate, but it cannot react when available capacity changes.

\subsection{ALVR built-in ABR}
ALVR's native adaptive mode, updates the encoder target bitrate once per second based on ALVR's internal capacity estimate. The estimate is derived from observed frame payload sizes and network delay and scaled by a conservative safety margin. Additional delay checks on the encoder, decoder, and network path can further reduce the bitrate when measured latency exceeds predefined thresholds. Overall, {ALVR's native ABR} follows a capacity‑driven bitrate adjustment strategy that relies on internal estimates and delay thresholds.

\subsection{Google Congestion Control}
Google Congestion Control (GCC)~\cite{carlucci2016analysis} is a delay‑based congestion control algorithm originally developed for WebRTC and widely used as a reference for low‑latency real‑time media transport. GCC estimates congestion from one‑way delay trends observed in receiver‑side arrival times rather than from explicit capacity probing and adjusts the sending rate accordingly. Conceptually, GCC tracks whether the network is in an overuse, underuse, or normal regime and increases the sending rate when congestion is not indicated, while decreasing it multiplicatively when overuse is detected. Thus, GCC serves as a representative general‑purpose transport‑level baseline in our comparison. 

\subsection{Network‑Assisted Dynamic Adaptation}
Network‑Assisted Dynamic Adaptation (NADA)~\cite{zhu2020network} (RFC 8698) is a unified congestion control scheme for interactive real‑time media. It combines delay, loss, and, when available, explicit congestion signals into a single receiver‑side congestion metric, which is then fed back to the sender to regulate the sending rate. NADA is designed to operate at the transport layer and to accommodate both implicit congestion cues (such as increased queuing delay and packet loss) and explicit signaling, serving thus as a general real‑time media baseline in our evaluation. 

\subsection{EVeREst-Intra}

EVeREst-Intra~\cite{korneev2024model}, hereinafter EVeREst, is a distributed client‑side bitrate adaptation algorithm for cloud VR streaming that operates without network assistance. 
It combines frame‑delivery delay monitoring with congestion estimation based on video-frames sizes to infer per‑stream sustainable link capacity and achievable throughput.
Based on these estimates and the current multi-user load, EVeREst computes a conservative bitrate bound with headroom for additional users and uses short-term and long-term delay thresholds to decide when to increase, decrease, or maintain the bitrate.

\paragraph*{Implementation details}  All baseline controllers are implemented in our open-source ALVR fork\footref{fn:github}. GCC and NADA are ported from Alhilal \textit{et al.}'s \cite{alhilal2025congestion} open-source ALVR implementation. EVeREst is implemented according to its published description, but assuming a single capacity and throughput estimation sample over the single ALVR video frame burst, instead of the maximum of samples over groups.

\section{Experimental Setup and Methodology} 



\subsection{In-Lab Testbed}\label{sec:setup_in}
The in-lab experiments are conducted in a controlled environment at the Department of Information and Communication Technologies (DTIC) of Universitat Pompeu Fabra, Barcelona, Spain.

\begin{figure}[t!!!]
    \centering
    \begin{subfigure}{\columnwidth}
        \centering
        \includegraphics[width=\linewidth]{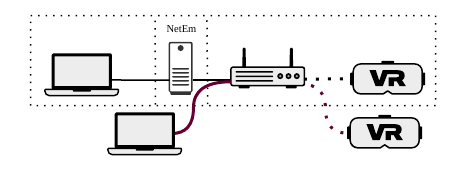}
        \caption{}
    \end{subfigure}
    \begin{subfigure}{\columnwidth}
        \centering
        \includegraphics[width=\linewidth]{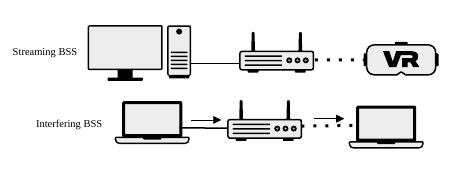}
        \caption{}
    \end{subfigure}
    \caption{Experimental in-lab testbed. (a) Single-/multi-user setup, with one or two active VR streaming pairs, respectively; \texttt{NetEm} is used only in bandwidth-constrained experiments. (b) Co-channel interference setup with two overlapping BSSs, where one BSS carries VR streaming traffic and the other generates background UDP traffic.}
    \label{Fig:testbed}
\end{figure}

The testbed consists of a wireless VR streaming basic service set~(BSS)\footnote{A BSS comprises a single access point and one or more associated stations.}, comprising a Wi-Fi access point~(AP) and one streaming server--HMD pair per user (see Fig.~\ref{Fig:testbed}). Depending on the scenario, the setup is extended with either an intermediate host for network emulation (a \texttt{NetEm}) or a second overlapping wireless network to generate background traffic. The former is used to evaluate controlled bandwidth-constrained scenarios, whereas the latter enables experiments under co-channel interference conditions. Table~\ref{tab:experiment_conditions} summarizes the in-lab equipment.

The \texttt{NetEm} is deployed between the streaming server and the AP and uses Linux traffic control (\texttt{tc})\footnote{\url{https://man7.org/linux/man-pages/man8/tc.8.html}} to emulate downlink bandwidth limitations, packet loss, and latency. The overlapping wireless network consists of a second AP and a sender--receiver pair generating downlink UDP traffic using \texttt{iperf3}\footnote{\url{https://iperf.fr}}.

Unless otherwise stated, no artificial fixed network delay is applied in our testbed. All servers, the \texttt{NetEm} (when used), and APs are interconnected via 1~Gbps Ethernet and are physically close, while HMDs connect wirelessly over Wi-Fi~6 in the 5~GHz band. All wireless links use two spatial streams, while OFDMA and MU-MIMO are disabled to avoid additional scheduling effects. Single-BSS experiments use an 80~MHz channel. For overlapping BSS~(OBSS) experiments, both APs operate on the same primary channel and channel bandwidth and with overlapping coverage areas, using either 80~MHz or 40~MHz depending on the scenario, thereby creating fully overlapping spectrum usage.

\subsection{Out-of-Lab Testbed}
The out-of-lab experiments, presented in Section~\ref{sec:crew}, are conducted at the facilities of CREW, an XR company specializing in immersive experiences, in Brussels, Belgium. The evaluation uses a professional XR environment following the same general streaming architecture as the in-lab testbed, but without network emulation or an additional overlapping BSS. Table~\ref{tab:crews_equipment} summarizes the out-of-lab equipment.

As in the in-lab setup, servers and the AP are connected through 1~Gbps Ethernet, whereas HMDs communicate wirelessly with the AP. The AP operates using an 80~MHz channel on either Wi-Fi~6 in the 5~GHz band or Wi-Fi~6E in the 6~GHz band. All wireless links use two spatial streams, with OFDMA and MU-MIMO disabled.

\begin{table}[t]
    \centering
    \caption{In-lab setup equipment details.}
    \begin{threeparttable}
    \scriptsize
\begin{tabular}{@{}p{\columnwidth/17*4} p{\columnwidth/8} p{\columnwidth/19*10}@{}}
\toprule
    \textbf{2x Laptop}\tnote{a} & Model & Dell G15 5521 \\
    & OS       & Windows 11 x64 \\
    & GPU  & NVIDIA GeForce RTX 3060 Mobile  \\
    & CPU                      &  i7-12700H  \\ 
    \textbf{1x PC}\tnote{b} & OS       & Windows 10 x64 \\
    & GPU  & NVIDIA GeForce RTX 3080, 10 GB  \\
    & CPU                      &  i5-12600KF  \\ 
    \midrule
    \textbf{1x PC} (\texttt{NetEm}) & Model & Dell Optiplex 3000 \\
      & OS       & Ubuntu Desktop 22.04.3 \\
    & GPU  & Intel UHD Graphics 770  \\
    & CPU                      &  i5-12500  \\ 
    \midrule
    \textbf{1x AP} (vr) & Model & ASUS ROG Rapture GT-AXE11000 \\
    &  Standard               & 802.11ax         \\
    \textbf{1x AP} (udp) & Model & ASUS TUF-AX4200 \\
    &  Firmware    &  OpenWrt 23.05.5 \\
    &  Standard               & 802.11ax         \\
    \midrule
    \textbf{2x HMD} & Model & Meta Quest 2 \\
     \bottomrule
    \end{tabular}
    \begin{tablenotes}[flushleft]
    \item[a] VR streamers except in the co-channel interference experiment, where they serve as endpoints for background traffic.
    \item[b] VR streamer only in the co-channel interference experiment.
\end{tablenotes}
    \end{threeparttable}
    \label{tab:experiment_conditions}
\end{table}

\begin{table}[t]
    \centering
    \caption{Out-of-lab setup equipment details.}
    \scriptsize
\begin{tabular}{@{} p{\columnwidth/17*4} p{\columnwidth/8} p{\columnwidth/19*10} @{}}
\toprule
     \textbf{2x PC } & OS       & Windows 11 x64 \\
    & GPU  & NVIDIA GeForce RTX 4090  \\
    & CPU                      &  i9-12900K \\
\midrule
    \textbf{1x AP} & Model & Netgear Nighthawk RAXE500 \\
    &  Standard               & 802.11ax         \\
    \midrule
    \textbf{2x HMD} & Model & HTC Vive Focus 3 \\
     \bottomrule
    \end{tabular}
    \label{tab:crews_equipment}
\end{table}



\subsection{Evaluation Scenarios}\label{sec:scenarios}
The in-lab testbed is used to benchmark NeSt-VR against the controllers introduced in Section~\ref{sec:controllers} under emulated network-capacity fluctuations, user mobility, and multi-user contention. It also evaluates NeSt-VR under co-channel interference. The out-of-lab testbed validates NeSt-VR in a professional XR environment. User mobility and multi-user contention experiments are repeated three times, and the reported results correspond to the mean across runs, with error bars indicating one standard deviation. The remaining experiments are performed once.
Table~\ref{tab:tests_config} summarizes the experimental configuration for each evaluation scenario.

\subsection{Streaming Configuration}
All experiments use our ALVR~v20.6.0 fork, which includes all evaluated bitrate management approaches. Default streaming settings are used, including a target frame rate of 90~FPS (72~FPS in the out-of-lab experiments). The in-lab experiments use \textit{SteamVR Home}, whereas the out-of-lab experiments use CREW's XR application \textit{Anxious Arrivals}\footnote{\url{https://crew.brussels/en/productions/anxious-arrivals}}. Unless otherwise stated, the bitrate bounds are 10--100~Mbps, with an initial bitrate of 100~Mbps in the in-lab experiments and 50~Mbps in the out-of-lab experiments. The constant-bitrate baseline is configured to 100~Mbps in the considered experiments. For ALVR ABR, the default parameters are used, including a saturation multiplier of 0.95 and latency thresholds for encoder, decoder, and network delays of 0.9 times the frame interval, 30 ms, and 8 ms, respectively. NeSt-VR uses the \textit{Balanced} profile by default, while the out-of-lab evaluation additionally considers the \textit{Speedy} and \textit{Anxious} profiles.

\subsection{NeSt-VR Settings}\label{sec:nestvr_settings}

The parameter values are listed in Table~\ref{tab:recommendations}. The control interval ($\tau=1$~s) and smoothing window ($t=1$~s) provide a good compromise between responsiveness and stability while filtering short-term network fluctuations. The capacity safety margin ($m=0.9$) limits the target bitrate to 90\% of the estimated capacity.
The operating thresholds are set to $\rho=0.99$ and $\sigma=22$~ms to maintain satisfactory user experiences at both 60~FPS and 90~FPS according to the in-lab perceptual tests reported in Table~\ref{tab:thresh_exp}. These thresholds are consistent with ITU-T recommendations~\cite{itu_t_y3109, itu_t_j1631}, which suggest round-trip times around 20~ms and packet loss below $10^{-5}$. 
Finally, the adaptation parameters are configured conservatively: $\gamma_{\text{rtt}}=1$ ensures that excessive delay always triggers a bitrate reduction, $\gamma_{+}=0.25$ limits upward bitrate exploration, $N_{\text{s}}=9$ defines the bitrate ladder granularity, and $N_{\text{up}}=1$ restricts bitrate increases to a single ladder step. The missing-feedback safeguard uses a warm-up period of $t_{\mathrm{warm}}=5$~s and is activated after $k_{\mathrm{miss}}=5$ consecutive missing feedback updates.

\subsection{Data Availability}
The dataset contains the ALVR streaming statistics collected during all experiments and is publicly available on Zenodo\footnote{\url{https://doi.org/10.5281/zenodo.14832268}\label{fn:zenodo}}.

\begin{table}[t!]
    \centering
    \caption{Subjective authors' assessment of average NFR under emulated packet-loss rates (1e-3--1e-6) and mean VF-RTT under added delays (6--40~ms).}
    \label{tab:thresh_exp}
    \scriptsize
    \begin{threeparttable}
    \begin{tabular}{@{}l@{\hspace{9pt}}ccc@{\hspace{5pt}}|@{\hspace{5pt}}l@{\hspace{9pt}}ccccc@{}}
        \toprule
        \textbf{NFR} & 0.91 & 0.99 & 0.999 & \textbf{VF-RTT} & 11 & 16 & 22 & 33 & 44 \\
        60 FPS & {\color{red}\ding{55}} & {\color{green}\ding{51}} & {\color{green}\ding{51}} & 60 FPS  & {\color{green}\ding{51}} & {\color{green}\ding{51}} & {\color{green}\ding{51}} & {\color{orange}\ding{55}} & {\color{red}\ding{55}}  \\
        90 FPS & {\color{red}\ding{55}} & {\color{green}\ding{51}} & {\color{green}\ding{51}} & 90 FPS & {\color{green}\ding{51}} & {\color{green}\ding{51}} & {\color{green}\ding{51}} & {\color{green}\ding{51}} & {\color{orange}\ding{55}} \\
        \bottomrule
    \end{tabular}   
    \begin{tablenotes}[flushleft]
        \item The values correspond to 30-second in-lab tests using CBR. The mean VF-RTT under nominal network conditions was 4--5~ms. 
        \textcolor{green}{\ding{51}} indicates no perceived visual impairments. \textcolor{red}{\ding{55}} (or \textcolor{orange}{\ding{55}}) indicates that all (or only some) evaluators perceived impairments. 
    \end{tablenotes}
    \end{threeparttable}
\end{table}

\begin{table}[t!]
\centering
\setlength{\tabcolsep}{3pt}
\caption{Experimental configurations for the in-lab and out-of-lab evaluations.}
\label{tab:tests_config}
\scriptsize
\begin{threeparttable}
\begin{tabular}{@{}l c c c c c c c c c l@{}}
\toprule
 \textbf{} & \textbf{(s)} & \textbf{\texttt{NetEm}} & \textbf{OBSS} & \textbf{C} & \textbf{A} & \textbf{G} & \textbf{D} & \textbf{E} & \textbf{N} \\
\midrule
 \ref{Sec:comparison_capacity} Capacity fluctuations & 80  & \checkmark & $\times$ & \checkmark & \checkmark & \checkmark & \checkmark & \checkmark & b \\
\ref{Sec:mobility_su}  Single-user mobility & 60  & $\times$ & $\times$ & \checkmark & \checkmark & \checkmark & \checkmark & \checkmark & b \\
 \ref{Sec:multiple_users} Multi-user contention & 60  & $\times$ & $\times$ & \checkmark & \checkmark & \checkmark & \checkmark & \checkmark & b\\
 \cmidrule(lr){1-10}
 \ref{Sec:obss}Co-channel interference & 120 & $\times$ & \checkmark & \checkmark & $\times$ & $\times$ & $\times$ & $\times$ & b \\
\midrule
 \ref{sec:crew} Out-of-lab deployment & 140 & $\times$ & $\times$ & \checkmark & $\times$ & $\times$ & $\times$ & $\times$ & b, s, x \\
\bottomrule
\end{tabular}
\begin{tablenotes}[flushleft]
\item \checkmark indicates that the corresponding configuration component or bitrate adaptation approach is used. 
\item C: CBR, A: \textit{Adaptive}, G: GCC, D: NADA, E: EVeREst, N: NeSt-VR (b: \textit{Balanced}, s: \textit{Speedy}, x: \textit{Anxious}).
\item 
\end{tablenotes}
\end{threeparttable}
\end{table}



\section{Results} \label{Sec:abr_evalu}
This section evaluates NeSt-VR under the scenarios introduced in Section~\ref{sec:scenarios} in terms of requested bitrate, reliability, and delay.

\subsection{Downlink Capacity fluctuations} \label{Sec:comparison_capacity}

\begin{table}[t!]
\centering
\caption{Streaming performance under network capacity fluctuations.}
\label{tab:capacity_fluctuations}
\setlength{\tabcolsep}{3pt}
\scriptsize
\begin{threeparttable}
\begin{tabular}{@{}lccccc@{}}
\toprule
\textbf{Controller} & \textbf{Av. (\%)} & \textbf{Th. (Mbps)} & \textbf{FPS} & \textbf{VF-RTT (ms)} & \textbf{PLR (\%)} \\
\midrule
CBR (100)     & 50.7 & 89.2 & 45.7 & 17.2 & 29.7 \\
ALVR ABR      & 73.1 & 66.2  & 64.3 & 15.7 & 19.6 \\
EVeREst       & 87.6 & 25.4  & 78.7 & 6.6  & 15.3 \\
GCC           & 87.4 & 24.9  & 77.0 & 32.8 & 4.7  \\
NADA          & 73.7 & 75.6  & 64.9 & 21.8 & 14.3 \\
\cmidrule(lr){1-6}
NeSt-VR       & 94.4 & 44.0  & 83.6 & 26.9 & 4.5  \\
\bottomrule
\end{tabular}
\begin{tablenotes}[flushleft]
\item \textit{Av.} denotes session availability, i.e., the percentage of the experiment during which the client remains connected. 
\textit{Th.} denotes the average delivered throughput during connected periods. \textit{FPS} denotes the mean session client frame rate. \textit{VF-RTT} corresponds to the mean VF-RTT over connected periods. \textit{PLR} denotes the packet loss rate.
\end{tablenotes}
\end{threeparttable}
\end{table}

This scenario evaluates controller behavior under controlled changes in the available downlink capacity, isolating the impact of bandwidth fluctuations on adaptation performance. A stepwise sequence of emulated downlink bandwidth limits is applied every 10~s: 150.0, 34.4, 103.2, 17.2, 137.6, 68.8, 51.6, and 150.0~Mbps. The lowest limits, 34.4 and 17.2~Mbps, represent the most severe drops, while 68.8 and 51.6~Mbps correspond to moderate capacity reductions. These values approximate Wi-Fi~6 data rates attainable with two spatial streams, a 20~MHz channel, and modulation and coding scheme~(MCS) levels~0--6, thereby covering both unconstrained and capacity-limited operating conditions. Throughout the experiment, the client remains approximately 1~m from the AP, with an average received signal strength indicator~(RSSI) of $-44$~dBm.

\begin{figure}[t!!]
  \centering
  \begin{subfigure}[b]{0.92\columnwidth}
    \includegraphics[width=\linewidth]{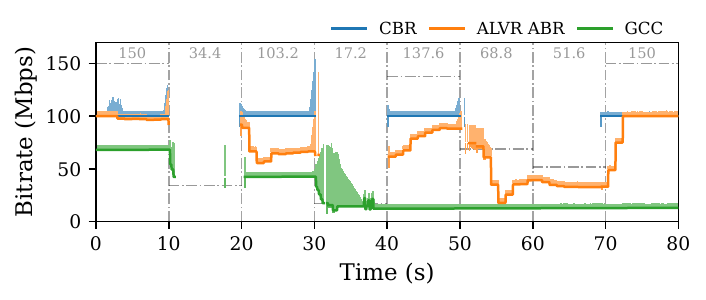}
    \caption{}
    \label{fig:}
  \end{subfigure}
  \begin{subfigure}[b]{0.92\columnwidth}
    \includegraphics[width=\linewidth]{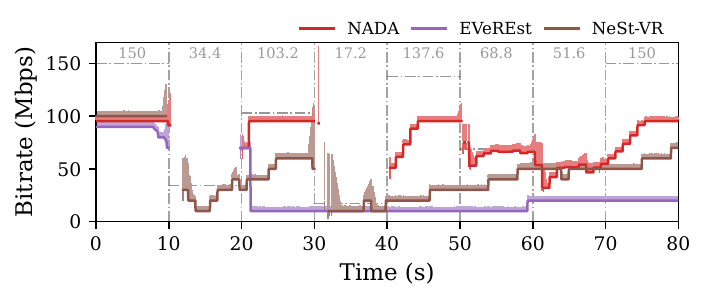}
    \caption{}
    \label{fig:}
  \end{subfigure}
    \caption{Temporal evolution of the requested bitrate under emulated network capacity fluctuations. (a) CBR at 100~Mbps, ALVR's native adaptive controller, and GCC. (b) NADA, EVeREst, and NeSt-VR. VF-RTT samples are shown above the bitrate curves, and frame-loss events are shown below them, with marker height proportional to event magnitude and reduced visual prominence relative to the bitrate traces. 
    }
  \label{fig:capacity_bitrate}
\end{figure}

Across all controllers, the most severe bandwidth reductions temporarily push the available capacity below the requested streaming bitrate, causing queue overflows and occasional disconnections.\footnote{Packet losses may still occur despite adaptation because removing the \texttt{tc} queueing discipline when updating the emulated bandwidth limit immediately drops any queued packets. Consequently, substantial packet and frame losses can occur even when the controller adapts the bitrate.}
Fig.~\ref{fig:capacity_bitrate} illustrates the temporal evolution of the requested bitrate, and Table~\ref{tab:capacity_fluctuations} summarizes the resulting streaming performance in terms of availability, delivered throughput during connected periods, session client frame rate, VF-RTT during connected periods, and packet loss.

CBR at 100~Mbps performs worst: it keeps the requested bitrate fixed at 100~Mbps throughout the experiment and thereby attains the lowest availability, the highest packet loss, and the lowest session client frame rate. During connected periods, it achieves an average throughput of 89.2~Mbps. Because it cannot react to capacity reductions, it repeatedly overloads the link, causing prolonged disconnections whenever the available capacity falls substantially below its fixed target.

ALVR's native adaptive controller similarly suffers from prolonged disconnections during the most severe intervals. Under moderate drops, it lowers the bitrate after experiencing high delay and packet loss, but it still remains well below the available capacity limit. During recovery intervals, it increases the bitrate relatively quickly. As a result, its average delivered throughput during connected periods remains high, but its session performance, in terms of frame rate and availability, remains limited.

EVeREst and GCC both favor conservative operation, but they also suffer disconnections during the first major congestion event. After that event, EVeREst settles to a near-minimum bitrate and remains close to that operating point for the rest of the trace. This protects delay, but it underutilizes the available capacity during recovery intervals and sacrifices visual quality. GCC behaves similarly, although its reaction is less abrupt: it reduces the bitrate after the first congestion event and then progressively lowers it during the subsequent severe drop until it settles near the minimum for the remainder of the experiment. In both cases, the controllers avoid sustained overload and achieve high availability, but fail to exploit the higher-capacity intervals effectively, resulting in substantially lower throughput.

NADA responds effectively to moderate capacity reductions but does not recover cleanly from abrupt collapses that temporarily interrupt the connection. When the reductions are moderate, it tracks the available capacity closely and keeps the requested bitrate near the instantaneous limit. In contrast, when the capacity drops are severe, it experiences complete disconnections. After recovery, it restores the requested bitrate quickly and reaches a high operating point again, indicating good performance during connected periods but lower robustness to abrupt severe limits. Accordingly, it achieves low availability while maintaining high throughput during connected periods.

NeSt-VR, in contrast, manages to recover the connection during the major drops while reducing the requested bitrate to remain near the instantaneous capacity limit. Under moderate drops, it converges to the highest bitrate attainable under the current bandwidth limit, and as conditions improve, it gradually probes for additional capacity. Thus, it increases the bitrate progressively, avoiding abrupt overshoot, but it does not reach the highest bitrate bounds during recovery periods. Accordingly, its average achieved throughput during connected periods is considerably lower than that of ALVR's adaptive controller and NADA, but substantially higher than that of GCC and EVeREst, while it achieves the highest availability, remaining connected for 94.4\% of the test. The higher availability directly translates into improved streaming performance, as it delivers the highest average client frame rate and the lowest packet loss rate.
In terms of delay, the mean VF-RTT during connected periods reaches 26.9~ms, exceeding the 22~ms satisfaction threshold. However, the median VF-RTT is only 9.7~ms, indicating that the controller remains within the desired operating region for most of the experiment while tracking the available capacity closely. The higher mean results from short latency spikes immediately after bandwidth reductions and reconnection, rather than from sustained delay inflation. These transient post-recovery delays trigger additional bitrate reductions before the controller progressively recovers, suggesting an opportunity to improve the controller's robustness.

These results show that, under capacity fluctuations, effective adaptation requires tracking the instantaneous capacity limit while still recovering quickly enough to avoid disconnections. NeSt-VR demonstrates effective adaptation by avoiding the persistent under-utilization of the most conservative controllers while providing a more stable operating point than the aggressive baselines, which suffer prolonged disconnections.

\subsection{Single-User Mobility}\label{Sec:mobility_su}

This scenario evaluates controller behavior under representative indoor mobility conditions, reflecting immersive VR applications in which users move within indoor environments while interacting with virtual content spatially aligned with their physical surroundings, rather than being limited to stationary usage or unrestricted movement in empty spaces. A single user (\textit{User~A}) moves between two locations while streaming VR content. Specifically, the user moves from Location~1 to Location~2 over 20~s, remains there for 20~s, and then returns to Location~1 over the final 20~s while walking at approximately 1.6~m/s. Location~1 is close to the AP, with an average RSSI of approximately $-48$~dBm, whereas Location~2 is farther away, with an average RSSI of approximately $-64$~dBm. This mobility introduces variations in the RSSI and, consequently, in the selected MCS and the achievable physical-layer data rate. Additionally, since walking induces significant head movements and alters the user's orientation with respect to the AP, it can increase Wi-Fi retransmissions in both the downlink and uplink, further reducing the achievable throughput~\cite{mehrdad2026breaking} and motivating bitrate adaptation. 

Across User~A's trajectory, all evaluated controllers keep the client frame rate close to the 90~FPS target, experience no disconnections, and maintain the mean VF-RTT well below the 22~ms satisfaction bound, with the highest average value remaining below 8.4~ms. 

\begin{figure}[t]
    \centering
    \begin{tikzpicture}
        \node[inner sep=0] (img) {
            \includegraphics[width=0.9\columnwidth]{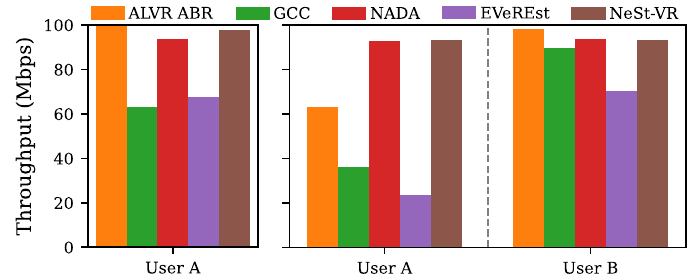}
        };
        \node at ([xshift=-57pt,yshift=-15pt]img.south) {\scriptsize{(a)}};
        \node at ([xshift=44pt,yshift=-15pt]img.south) {\scriptsize{(b)}};
    \end{tikzpicture}
    \caption{Average achieved throughput under each bitrate controller. (a) Single-user mobility scenario. (b) Multi-user contention scenario.}
    \label{fig:mobility_bitrate}
\end{figure}

Fig.~\ref{fig:mobility_bitrate}a shows the average achieved throughput across trials. NeSt-VR maintains the requested bitrate near the upper end of the operating range throughout the experiment, achieving an average throughput of 97.8~Mbps. 
ALVR's native adaptive controller maintains a similar throughput level, reaching 100~Mbps on average. However, unlike NeSt-VR, it incurs a non-zero packet-loss rate ($2.9\times10^{-5}$), reflecting its delay-based adaptation, which does not explicitly account for reliability.
NADA also maintains a high operating point, achieving an average throughput of 93.8~Mbps while exhibiting negligible frame loss.
In contrast, GCC and EVeREst operate at substantially lower average bitrates, achieving 63.2~Mbps and 68.0~Mbps average throughputs, respectively, representing reductions of 35.4\% and 30.5\% compared with NeSt-VR. Because the available network capacity remains sufficient throughout the experiment, this added conservatism does not yield measurable gains in delay or reliability. Instead, it mainly reflects a more cautious response to RSSI variation than the channel conditions require.

These results are consistent with the findings of \cite{michaelides2025lessons}, which showed that an 80~MHz Wi-Fi~6 link can sustain high-quality VR streaming at RSSI values down to approximately $-64$~dBm. Accordingly, NeSt-VR maintains a near-maximum bitrate while preserving low delay and high reliability, whereas GCC and EVeREst adopt more conservative operating points than required under these mild mobility conditions. The next experiment introduces a second stationary user at the weaker location to create network contention and assess controller behavior under a more demanding operating regime.

\subsection{Multi-user contention}\label{Sec:multiple_users}

This scenario evaluates controller behavior under concurrent VR streaming, where two users share the same Wi-Fi BSS and contend for the wireless medium using the same bitrate adaptation controller. User~A follows the mobility trajectory described in Section~\ref{Sec:mobility_su}, whereas User~B remains stationary at Location~2.
Because User~B operates from a weaker radio location, its lower PHY transmission rate increases airtime consumption and reduces the transmission opportunities available to User~A. This behavior corresponds to the IEEE~802.11 \emph{performance anomaly}~\cite{heusse2003performance} and has also been observed in wireless VR streaming scenarios~\cite{michaelides2025lessons}.

Across all evaluated controllers, both users maintain frame rates close to the target 90~FPS without disconnections, while the mean VF-RTT remains comfortably below the 22~ms operating threshold. In particular, the highest mean VF-RTT observed is only 10.4~ms. 
As illustrated in Fig.~\ref{fig:mobility_bitrate}b, NeSt-VR maintains a high operating point for both users, achieving average throughputs of 93.4~Mbps and 93.1~Mbps for Users~A and~B, respectively, while maintaining negligible packet loss. NADA achieves similar throughputs, 92.7~Mbps and 93.6~Mbps, respectively, although it exhibits significantly higher packet loss for User~A ($1.01\times10^{-3}$). NeSt-VR also maintains bounded bitrate variability under contention, with a lower bitrate-change rate than NADA for the moving user, while keeping the stationary user's bitrate stable. ALVR's native adaptive controller behaves more asymmetrically, achieving  63.1~Mbps for the moving User~A and 98.3~Mbps for User~B. These values correspond to a 32.4\% lower throughput than NeSt-VR for User~A and a 5.6\% higher throughput for User~B, respectively, suggesting a stronger reaction to mobility-induced channel variations.
In contrast, as in the single-user scenario, GCC and EVeREst remain substantially more conservative, particularly for User~A, achieving only 36.3~Mbps and 22.5~Mbps, respectively. These values correspond to reductions of 61.1\% and 75.9\% relative to NeSt-VR. These reductions do not provide measurable improvements in reliability or delay, indicating that both controllers react conservatively to the combined effects of mobility and contention.

Overall, the results indicate that the available Wi-Fi capacity remains sufficient to sustain both streams under the considered conditions. Accordingly, NeSt-VR successfully maintains a near-maximum operating point while keeping delay low and reliability high. In contrast, more conservative baselines sacrifice bitrate without providing significant improvements in reliability or latency.

\paragraph{Case study: operation under severe channel degradation}

To further examine controller behavior under capacity-limited conditions, we consider a stress case in which User~B remains at a more challenging position with an average RSSI of approximately $-78$~dBm. The reduced signal strength lowers the achievable PHY transmission rate of the distant user, increasing airtime consumption and intensifying contention within the BSS. Here, a constant-bitrate baseline with a target bitrate of 100~Mbps serves as the comparative reference, representing an aggressive fixed-rate streaming configuration that stresses the wireless link.

Under these conditions, CBR experiences substantial degradation, particularly when both users operate at the weaker location and their MCSs, and consequently their available transmission rates, decrease. In particular, User~A suffers from occasional disconnections, severe latency spikes, and increased packet losses. As a result, the constant-bitrate baseline achieves mean VF-RTTs of 83.7~ms and 76.0~ms, packet-loss rates of 0.32\% and 0.16\%, and frame rates of 83.7~FPS and 86.6~FPS for Users~A and~B, respectively.
In contrast, NeSt-VR dynamically adapts the requested bitrate to the available wireless capacity, avoiding the excessive queueing and packet losses observed with the fixed-rate configuration. It maintains mean VF-RTTs of 13.1~ms and 17.2~ms for Users~A and~B, respectively, while sustaining approximately 90~FPS and eliminating packet losses. User~B converges to a lower target bitrate than User~A due to its consistently weaker channel conditions, requesting 42.3~Mbps in average compared with 81.1~Mbps for User~A. Hence, it reduces the mean VF-RTT by 84.3\% and 77.4\% for Users~A and~B relative to CBR, at the expense of reducing the target bitrate by only 18.9\% and 57.7\%, respectively.

These results highlight the benefit of bitrate adaptation under highly variable wireless conditions: instead of maintaining an unsustainable fixed bitrate, NeSt-VR adjusts the streaming rate to the available capacity and preserves the latency and reliability levels required for satisfactory VR streaming.


\begin{figure}[t] 
     \centering
     \begin{minipage}[b]{\columnwidth}
        \centering
        \includegraphics[width=\textwidth]{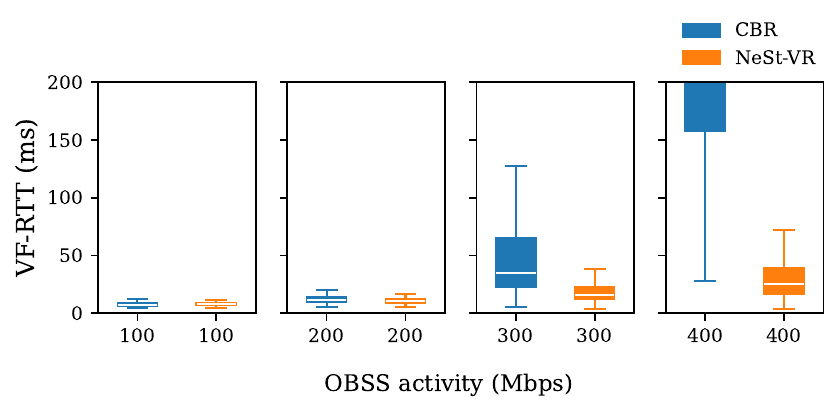}
        \caption{VR streaming VF-RTT distributions at 40~MHz under distinct OBSS activity levels. \normalfont{Note that with CBR and a 400~Mbps OBSS activity, the third quartile (75\%) exceeds 400~ms.} }
        \label{fig:vf_rtt_boxes_obss_40MHz}
    \end{minipage}
\end{figure}

\begin{figure}[t] 
     \centering
     \begin{minipage}[b]{\columnwidth}
        \centering
        \includegraphics[width=\textwidth]{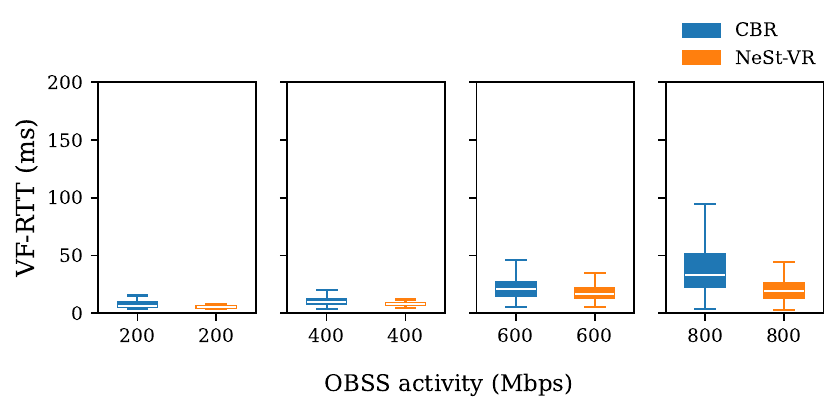}
        \caption{VR streaming VF-RTT distributions at 80~MHz under distinct OBSS activity levels.}
        \label{fig:vf_rtt_boxes_obss_80MHz}
    \end{minipage}
\end{figure}

\subsection{Co-channel interference} \label{Sec:obss}

Overlapping basic service sets (OBSSs) are common in dense Wi-Fi deployments, where multiple APs operate on overlapping channels and contend for medium access. Since Wi-Fi stations defer transmissions when the channel is sensed busy, OBSS traffic increases channel-access delays and reduces the airtime available to each BSS~\cite{haxhibeqiri2024coordinated, zhong2015issues}. Although Wi-Fi 6 introduces spatial reuse mechanisms, such as BSS Coloring and OBSS Preamble Detection, to mitigate interference from neighboring BSSs, overlapping deployments can still result in increased contention and reduced airtime availability, thereby degrading delay-sensitive applications such as VR streaming. This section evaluates the impact of OBSS traffic on VR streaming performance and compares bitrate adaptation algorithm against a constant-bitrate baseline.

As described in Section~\ref{sec:setup_in}, during VR streaming, a neighboring BSS generates background UDP traffic on the same channel. The VR client remains approximately 1~m from the AP, with an average RSSI of $-45$~dBm. The considered OBSS load levels are 100, 200, 300, and 400~Mbps at 40~MHz, and 200, 400, 600, and 800~Mbps at 80~MHz. At 40~MHz, CBR no longer operates beyond 400~Mbps of OBSS traffic.
Figures~\ref{fig:vf_rtt_boxes_obss_40MHz} and~\ref{fig:vf_rtt_boxes_obss_80MHz} summarize the VF-RTT distributions under the considered OBSS loads.

At 40~MHz, the lower OBSS loads of 100 and 200~Mbps are accommodated by both controllers without requiring bitrate adaptation. At 300~Mbps, however, co-channel contention becomes more pronounced. CBR experiences sporadic packet loss (0.28\%) and a mean VF-RTT of 54.6~ms, whereas NeSt-VR reduces the requested bitrate to 55.0~Mbps, maintaining a mean VF-RTT of 18.7~ms and avoiding packet loss. At 400~Mbps, contention further degrades CBR performance, causing sustained packet loss (15.64\%), reducing the delivered frame rate to 72.5~FPS, and producing occasional disconnections (99.3\% session availability). The resulting mean VF-RTT during connected periods reaches 227.0~ms. In contrast, NeSt-VR reduces the requested bitrate to its minimum value (10~Mbps), eliminates packet loss, and limits the mean VF-RTT to 30.4~ms, which remains within the 33~ms satisfaction range associated with 90~FPS operation.

At 80~MHz, the wider channel provides additional headroom, and no bitrate adaptation is required up to 600~Mbps. At 800~Mbps, although CBR experiences no packet loss, its mean VF-RTT increases to 42.4~ms, exceeding the satisfaction threshold. NeSt-VR, in contrast, adjusts the requested bitrate to 35.1~Mbps on average, maintaining a mean VF-RTT of 21.5~ms and providing a clear delay advantage under heavy OBSS contention.

Overall, the results show that OBSS activity can significantly degrade VR streaming, particularly at 40~MHz, with performance becoming increasingly vulnerable as channel contention rises. An effective adaptation algorithm such as NeSt-VR adjusts the streaming bitrate to the available network conditions, reducing traffic load and mitigating the impact of such contention.


\begin{figure}[t] 
     \centering
     \begin{minipage}[b]{\columnwidth}
        \centering
        \includegraphics[width=0.565\textwidth]{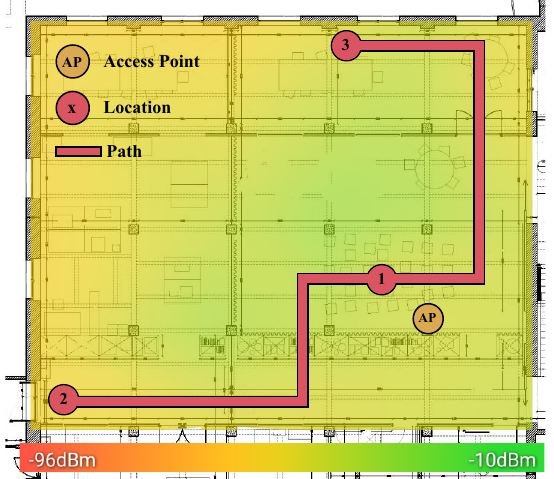}
        \caption{CREW facilities layout and RSSI heatmap, illustrating the indoor environment and the mobility end-points considered in the evaluation. 
        }
        \label{fig:crew_floor_plan}
    \end{minipage}
\end{figure}

\subsection{NeSt-VR in a professional XR venue}\label{sec:crew}

To evaluate the proposed controller in a realistic professional XR environment, we conducted experiments in CREW's facilities\footnote{\url{https://crew.brussels/}}, an indoor space of approximately ${\scriptscriptstyle \sim}12.5\times10$~m, using their XR application \textit{Anxious Arrivals}. CREW develops immersive XR performances that allow audiences to engage with alternate realities and interact in the (not) here and (not) now, blending physical and virtual presence. This provides a representative deployment scenario beyond controlled laboratory conditions.

\begin{figure}[t!!!]
    \centering
    \begin{subfigure}{\columnwidth}
        \centering
        \includegraphics[width=\linewidth]{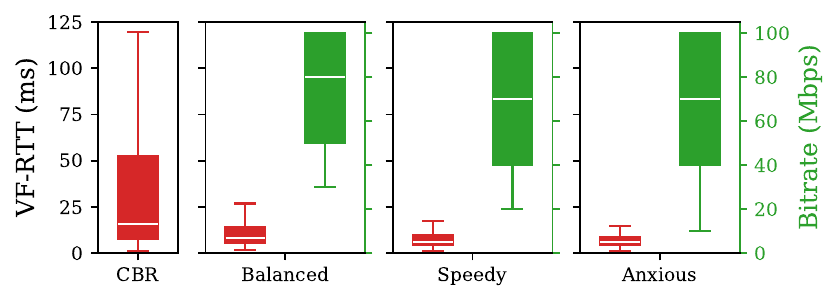}
    \end{subfigure}
    \caption{CREW multi-user evaluation: VF-RTT and requested bitrate distributions for \textit{User~C} under CBR and the NeSt-VR \textit{Balanced}, \textit{Speedy}, and \textit{Anxious} profiles.}
    \label{fig:crew_performance}
\end{figure}

The evaluation compares NeSt-VR's bitrate adaptation profiles against a constant-bitrate baseline in a multi-user mobility scenario involving \textit{User~C} and \textit{User~D}, who stream concurrently. The users move between the AP-adjacent \textit{Location~1} and the more distant \textit{Locations~2} and~\textit{3}, as illustrated in Fig.~\ref{fig:crew_floor_plan}. Since both users exhibit similar behavior, Fig.~\ref{fig:crew_performance} reports the results for \textit{User~C} only. 

Among the profiles, \textit{Balanced} achieves the highest average requested bitrate (74.5~Mbps) and a mean VF-RTT of 17.8~ms, reflecting its more conservative adaptation strategy. \textit{Speedy} decreases the bitrate faster under performance degradation, reducing the mean VF-RTT to 12.9~ms at the cost of a lower average requested bitrate (67.1~Mbps). \textit{Anxious} provides the lowest mean VF-RTT (8.1~ms) by applying the most aggressive bitrate reductions under degraded conditions. Nevertheless, it maintains an average requested bitrate of 68.7~Mbps, slightly higher than \textit{Speedy}. These results demonstrate that the profiles offer different latency–quality trade-offs, enabling adaptation behavior to be tailored to the latency sensitivity and visual quality requirements of the XR application.

Across all profiles, bitrate reductions are triggered when users move towards the more distant locations. As the distance from the AP increases, path loss reduces the received signal strength and the achievable PHY rate, lowering the effective throughput available to the VR streams. NeSt-VR adapts the requested bitrate accordingly, maintaining stable low-delay operation despite mobility, physical obstructions, and uncontrolled wireless conditions in a professional XR venue.

\begin{figure}[t] 
     \centering
     \begin{minipage}[b]{\columnwidth}
        \centering
        \includegraphics[width=\textwidth]{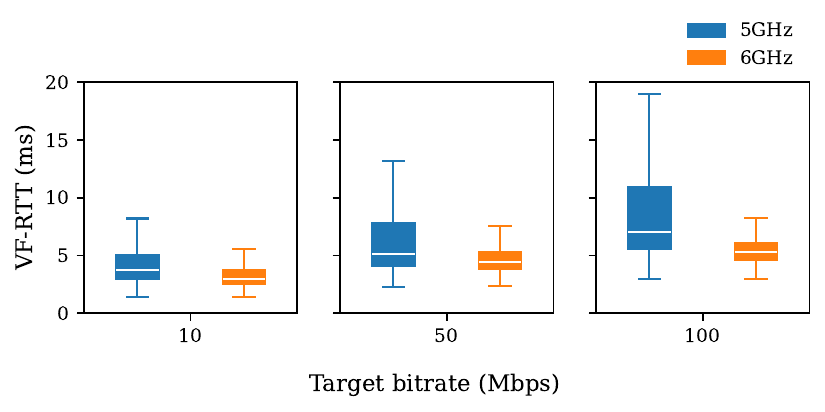}
    \caption{CREW single-user evaluation across 5~GHz and 6~GHz bands: VF-RTT distributions for CBR streaming at 10, 50, and 100~Mbps.}
    \label{fig:5ghz_6ghz}
    \end{minipage}
\end{figure}

\paragraph{Case study: 5~GHz vs. 6~GHz}

In addition to the multi-user experiments, we compared the 5~GHz and 6~GHz bands using single-user CBR streaming at 10, 50, and 100~Mbps. The corresponding VF-RTT distributions are shown in Fig.~\ref{fig:5ghz_6ghz}. The 6~GHz band consistently provides lower delays, particularly at higher bitrates, benefiting from additional spectrum availability and typically lower contention. Moreover, VF-RTTs increase rapidly with bitrate in the 5~GHz band, whereas the increase is considerably slower in the 6~GHz band, indicating greater headroom for high-bitrate XR streaming. These results further support the suitability of the 6~GHz band for demanding wireless XR applications.


\section{Conclusions}

This paper presented NeSt-VR, a network-aware step-wise adaptive bitrate controller for interactive VR streaming that adjusts the requested bitrate under dynamic Wi-Fi conditions using video-frame delivery and frame-delay feedback. Across representative Wi-Fi~6 scenarios, including user mobility, multi-user contention, fluctuating network capacity, and overlapping basic service sets, NeSt-VR consistently maintained low video-frame latency, high frame-delivery performance, and high visual quality, thereby satisfying the stringent latency and reliability requirements of immersive XR applications. In most scenarios, the resulting VF-RTT remained below 22~ms, which we identified as sufficient to ensure satisfactory user experience at both 60~FPS and 90~FPS. Compared with state-of-the-art controllers, including GCC, NADA, and EVeREst, it sustained higher bitrates without compromising reliability or latency. Under capacity fluctuations, it achieved the highest session availability (94.4\%) and frame rate (83.6~FPS), outperforming the best baseline by 6.8 percentage points and 4.9~FPS, respectively. Under mobility and multi-user contention, it sustained near-maximum throughput while preserving low delay and high reliability, whereas baselines such as GCC and EVeREst adopted more conservative operating points, resulting in significant throughput reductions (e.g., 35.4\% and 30.5\% under mobility, respectively). Under co-channel interference, it preserved low-delay operation and avoided the disconnections observed with fixed-rate streaming, reducing delays by up to 65.7\% in the most demanding conditions. These results were also consistent with its effectiveness in a professional XR venue under uncontrolled wireless conditions.

The open-source implementation released with this work enables reproducible evaluation and further research on adaptive VR streaming over Wi-Fi. Future work will investigate fairness-aware adaptation for dense multi-user XR environments, online learning-based parameter tuning mechanisms, trace-driven evaluation under diverse real-world network conditions, and objective visual-quality assessment. We also plan to incorporate access-point assistance and additional Wi-Fi feedback to further enhance controller operation~\cite{casasnovas2026bravr}, and to explore the interaction of NeSt-VR with emerging Wi-Fi mechanisms such as multi-link operation and multi-AP coordination~\cite{galati2024will}.


\section{Acknowledgment}

This work was supported by the following projects: MAX-R (GA 101070072 European Union), Wi-XR PID2021-{\allowbreak}123995NB-{\allowbreak}I00 and TRUE Wi-Fi PID2024-{\allowbreak}155470NB-{\allowbreak}I00 (MICIU/{\allowbreak}AEI/{\allowbreak}10,13039/{\allowbreak}501100011033/{\allowbreak}FEDER,UE), REALM (GA 101298050 European Union), 
ICREA Academia 2024 (00077 AGAUR), and MdM-UoE CEX2021-{\allowbreak}001195-{\allowbreak}M (MICIU/{\allowbreak}AEI/{\allowbreak}10.13039/{\allowbreak}501100011033).

\bibliographystyle{unsrt} 
\bibliography{bib}

\end{document}